\documentclass[12pt,aps,prb,preprint]{revtex4}

\usepackage{amsmath}
\usepackage{graphicx}
\usepackage{graphics}
\usepackage{tabularx}

\begin{document}

\title{How does a drinking straw balance work?}
\author{Hyunbyuk Kim}
\email{time@kaist.ac.kr}
\author{Jungwoo Park}
\affiliation{Department of Physics and Earth Science, Korea Science Academy of KAIST, Busan 614-822, South Korea}

\begin{abstract}
We examine working mechanism of a drinking straw balance. Unlike the usual horizontal level balances which rely on the pivotal rotation, a drinking straw balance shows rolling motion. Because of this rolling and relative position between the pin and the center line of a straw, a new equilibrium state is possible after placement of a test mass. Relative errors between the measured values by a drinking straw balance and an electronic scale are order of 22\%. Minimum value of a measurable mass by a straw balance is order of 10 mg.
\end{abstract}

\maketitle

\section{Introduction}
A drinking straw balance is a useful demonstration kit for the introductory physics.$^{\cite{bib:pssc},\cite{bib:unesco},\cite{bib:edge}}$ It is capable of measuring a tiny mass relatively precisely compared to its simplicity. There are literatures which introduce this material but as far as authors know there is no literature which treats the working mechanism behind its utility. One may think that the mechanism might be nothing but the fact that the mass ratio is an inverse of arm length ratio. Truly, many mass balances are working on this basic fact. One example is a meter stick balance which is a kind of horizontal level balance. A meter stick can rotates pivotally with its axis position fixed and horizontal level is set by adjusting the arm length of the known mass. Then the unknown mass can be inferred from the known mass and the arm length ratio. If we use a drinking straw balance in this way of horizontal level balances, its working mechanism is exactly same with theirs. But if we use a drinking straw balance as a mass indicating system then its mechanism requires more involved calculations. The essential difference between a drinking straw balance and a horizontal level balance is this. A drinking straw balance does not show pivotal rotation but does show rolling motion of the axis.

We study the theory of this rolling motion and the equilibrium first. Then we show the experimental results.

\section{Theory}
A drinking straw attached with a known mass at one end is prepared. A pin is piercing the straw at a proper point which is closer to the known mass side. This point is determined such a way that the drinking straw will sustain horizontal level when it is placed on the glass support. Now we place some tiny test mass on the other end of a straw. The straw starts to tilt and is no more horizontal. Our question is what is the mass of a test material?

Before placement of a test mass the straw was already in a horizontal equilibrium. So the test mass might break the equilibrium and a straw maybe start to tilt without bound. But actually it settles down to a new equilibrium. This is because the pin which is the axis of a balance is rolling not rotating. Due to this rolling motion straw's relative lever arm length can change to reach a new equilibrium.

\begin{figure}[h]
\begin{center}
\includegraphics[trim = 10mm 50mm 40mm 100mm, clip, width=1.0\textwidth]{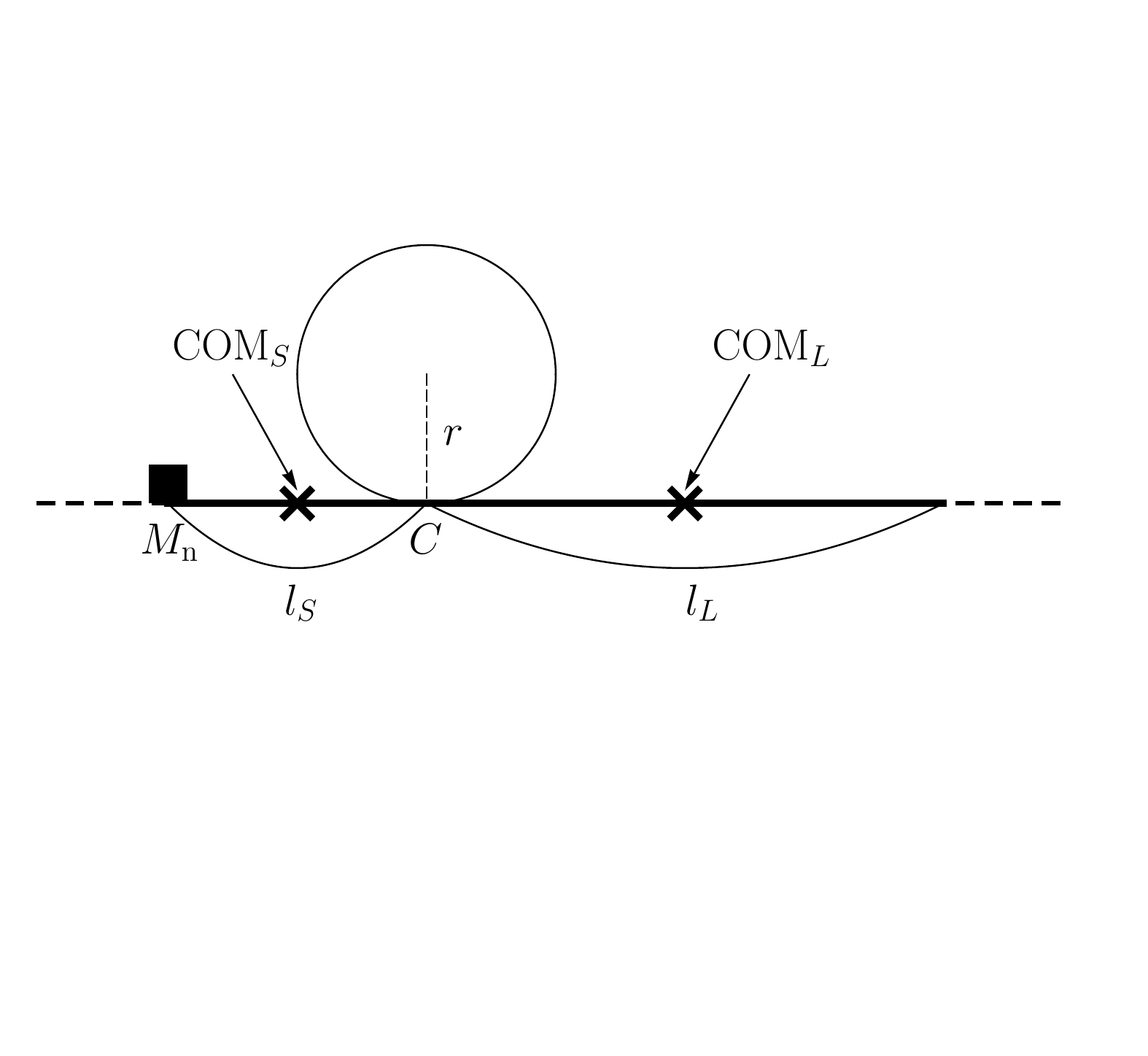}
\caption{\label{fig:horz}Initial equilibrium of a drinking straw balance. It is in a horizontal state with an attached mass. Solid line is a center line of a straw. Center of the circle is the center of a pin. We use a small nail for a known mass $M_{\rm n}$. Total mass of a straw is $M_{\rm st}$. ${\rm COM}_S$ and ${\rm COM}_L$ are centers of masses of short arm and long arm respectively. $l_S$ and $l_L$ are lengths of short arm and long arm respectively. $C$ is a supporting point. $r$ is a distance between the center of a pin and the center line of a straw.}
\end{center}
\end{figure}

The scheme of the equipment is given at Fig.~\ref{fig:horz}. The solid line represents the center line of a straw. The circle is drawn for clear demonstration of the rolling motion of a pin. The radius of the circle is a distance between the center of a pin and the center line of a straw. This distance is not the radius of a pin. Note that the center line of a straw is beneath the center of a pin, i.e., the pin is piercing the straw above the center line of it. The reason of this will become clear shortly. 

At horizontal equilibrium, the torque to the point $C$ due to the weight of $M_{\rm n}$ and the weight of left part of a straw must be same with the torque due to the weight of right part of a straw.$^{\cite{bib:HRW}}$
\begin{equation}
\label{eq:eqh}
l_S \times M_{\rm n} + \frac{l_S}{2} \times \left( \frac{l_S}{l_S+l_L} M_{\rm st} \right) = \frac{l_L}{2} \times \left( \frac{l_L}{l_S+l_L} M_{\rm st} \right).
\end{equation}
In this equation we assumed uniform mass density of a straw and canceled the gravitational acceleration.

\begin{figure}[h]
\begin{center}
\includegraphics[trim = 10mm 50mm 40mm 100mm, clip, width=1.0\textwidth]{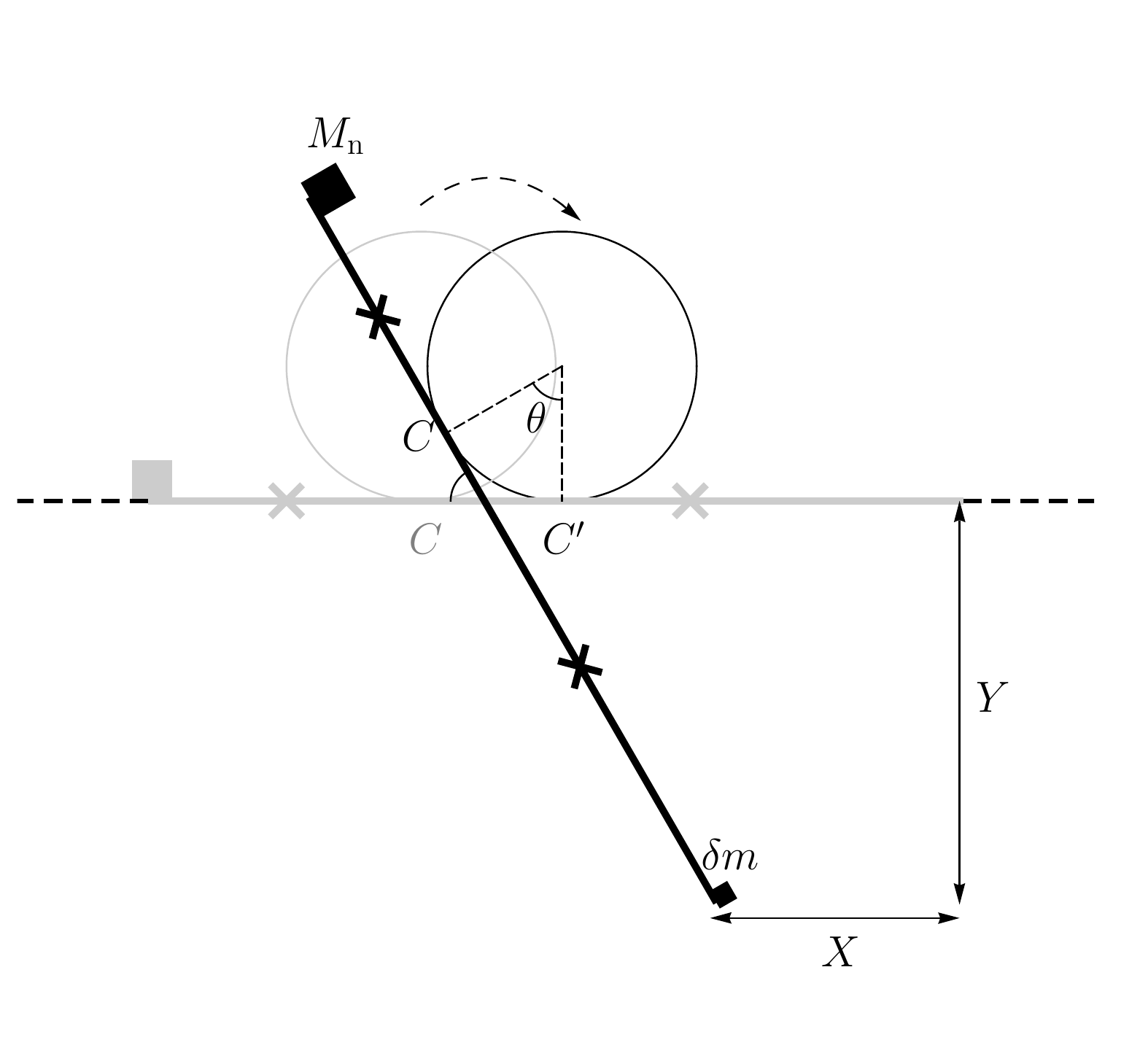}
\caption{\label{fig:tilt}New equilibrium is established after placement of $\delta m$. $C'$ is a new supporting point. $X$ and $Y$ are horizontal and vertical displacements of the end point respectively.}
\end{center}
\end{figure}

When we place a test mass $\delta m$ on the right end of a straw, the straw establishes a new equilibrium state with a tilted angle $\theta$ (see Fig.~\ref{fig:tilt}). Now the torques should be considered about a point $C'$. Note this is not the point $C$. New balance between the torques gives another equation. 
\begin{gather}
\label{eq:eqt}
\left( l_S \cos \theta + r \sin \theta \right) \times M_{\rm n} + \left( \frac{l_S}{2} \cos \theta + r \sin \theta \right) \times \left( \frac{l_S}{l_S+l_L} M_{\rm st} \right)\\ 
= \left( l_L \cos \theta - r \sin \theta \right) \times \delta m + \left( \frac{l_L}{2} \cos \theta - r \sin \theta \right) \times \left( \frac{l_L}{l_S+l_L} M_{\rm st} \right).
\nonumber
\end{gather}

From Eqs. (\ref{eq:eqh}) and (\ref{eq:eqt}) we can derive the formula for a test mass.
\begin{equation}
\label{eq:dmexac}
\delta m = \frac{r \sin \theta}{l_L \cos \theta - r \sin \theta} \left( M_{\rm st} + M_{\rm n} \right).
\end{equation}

Here are some points worth mentioning. First, the relative position of the center line of a straw and the center of mass (c.o.m) of a pin. If the position of the center line of a straw is above the c.o.m of a pin, the sign of $r$ in Eqs.~(\ref{eq:eqt}) and (\ref{eq:dmexac}) becomes opposite and there cannot be a new equilibrium state. In other words if the pin is piercing under the center line of a straw, this straw cannot function as a balance and even worse it becomes very hard to prepare a horizontal equilibrium from the beginning. Higher position of the c.o.m of a pin relative to the center line of a straw is one of two reasons why the straw balance works. Second, due to the rolling of a pin the relative length of lever arms is changed. If there is only a rotation this ratio is kept to be same and a new equilibrium state will never be established. Because of these two reasons the ratio of right lever arm length over left lever arm length becomes smaller after the straw tilts so that a new equilibrium state becomes possible. Third, from Eq.~(\ref{eq:dmexac}) we can see that if we use a smaller straw mass $M_{\rm st}$ and a smaller known mass $M_{\rm n}$ we can measure a tinier mass. Also with a longer straw we can measure a tinier mass.

Equation~(\ref{eq:dmexac}) is inconvenient to use because direct measurement of $\theta$ can induce large error. Instead we deduce $\theta$ from parameters which can be measured more precisely. From the geometry of Fig.~\ref{fig:tilt}, we have
\begin{eqnarray}
\label{eq:XY1}
X & = & l_L \left( 1 - \cos \theta \right) - r \left( \theta - \sin \theta \right).\\
\label{eq:XY2}
Y & = & l_L \sin \theta - r \left( 1 - \cos \theta \right).
\end{eqnarray}

By using these equations we can find $\theta$ numerically from the measured values of $X, Y, l_L$ and $r$. Two $\theta$ values from Eqs.~(\ref{eq:XY1}) and (\ref{eq:XY2}) give us additional check for consistency.

From Eqs.~(\ref{eq:dmexac}), (\ref{eq:XY1}) and (\ref{eq:XY2}) we can roughly estimate $\delta m$ by using the fact $l_L \gg r$.
\begin{equation}
\label{eq:dmapp}
\delta m \approx \frac{r}{l_L} \frac{Y}{l_L - X} \left( M_{\rm st} + M_{\rm n} \right).
\end{equation}

As we shall see $r/l_L \sim O(10^{-2})$, $Y/(l_L-X) \sim O(1)$ and $M_{\rm st}+M_{\rm n} \sim O(1\,{\rm g})$. So we can measure $\delta m$ of order 10 mg.

\section{Experiment}
\begin{figure}[h]
\begin{center}
\includegraphics[width=0.9\textwidth]{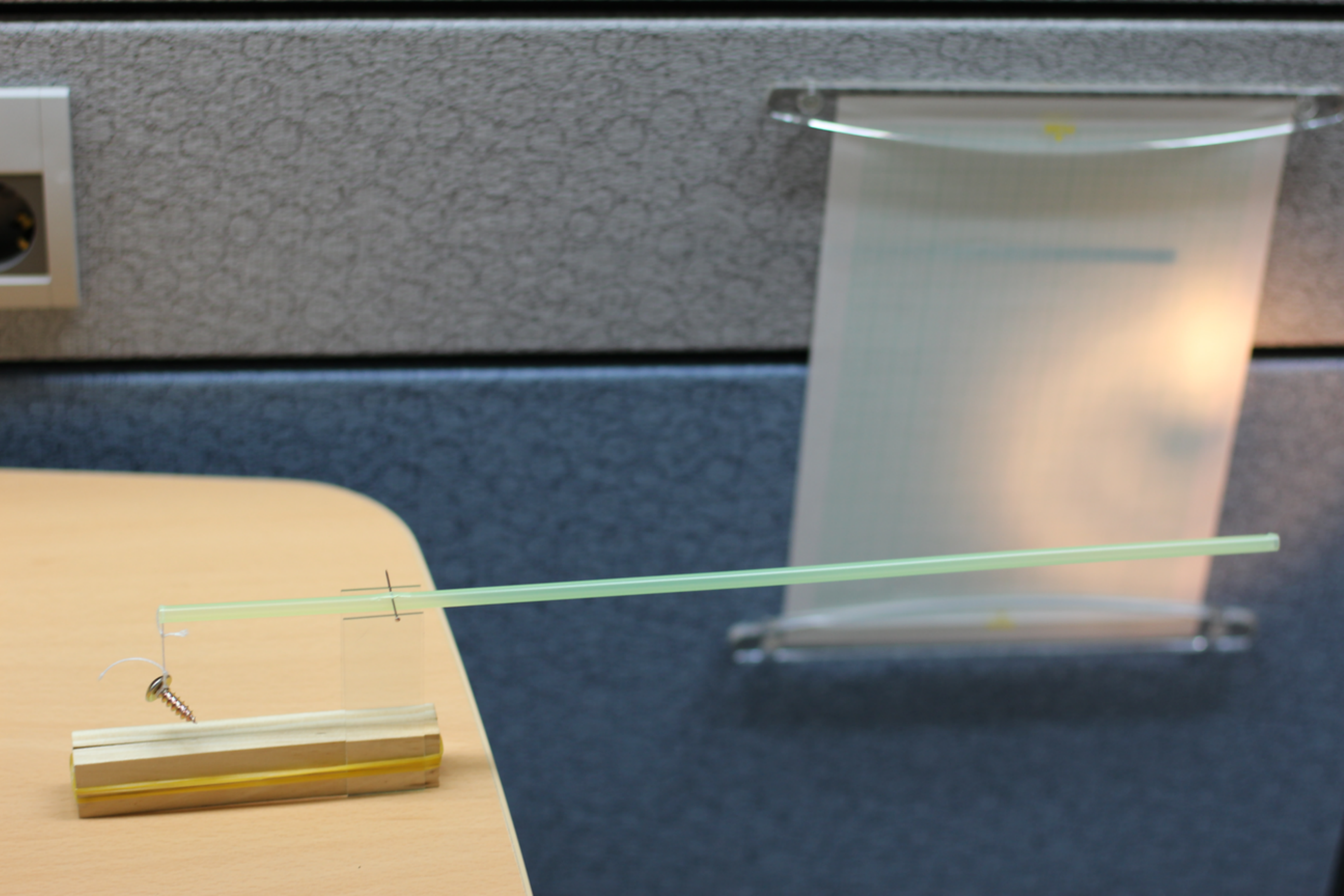}
\caption{\label{fig:exph}A drinking straw is in a horizontal equilibrium initially with a nail at one end.}
\end{center}
\end{figure}

\begin{figure}[h]
\begin{center}
\includegraphics[width=0.9\textwidth]{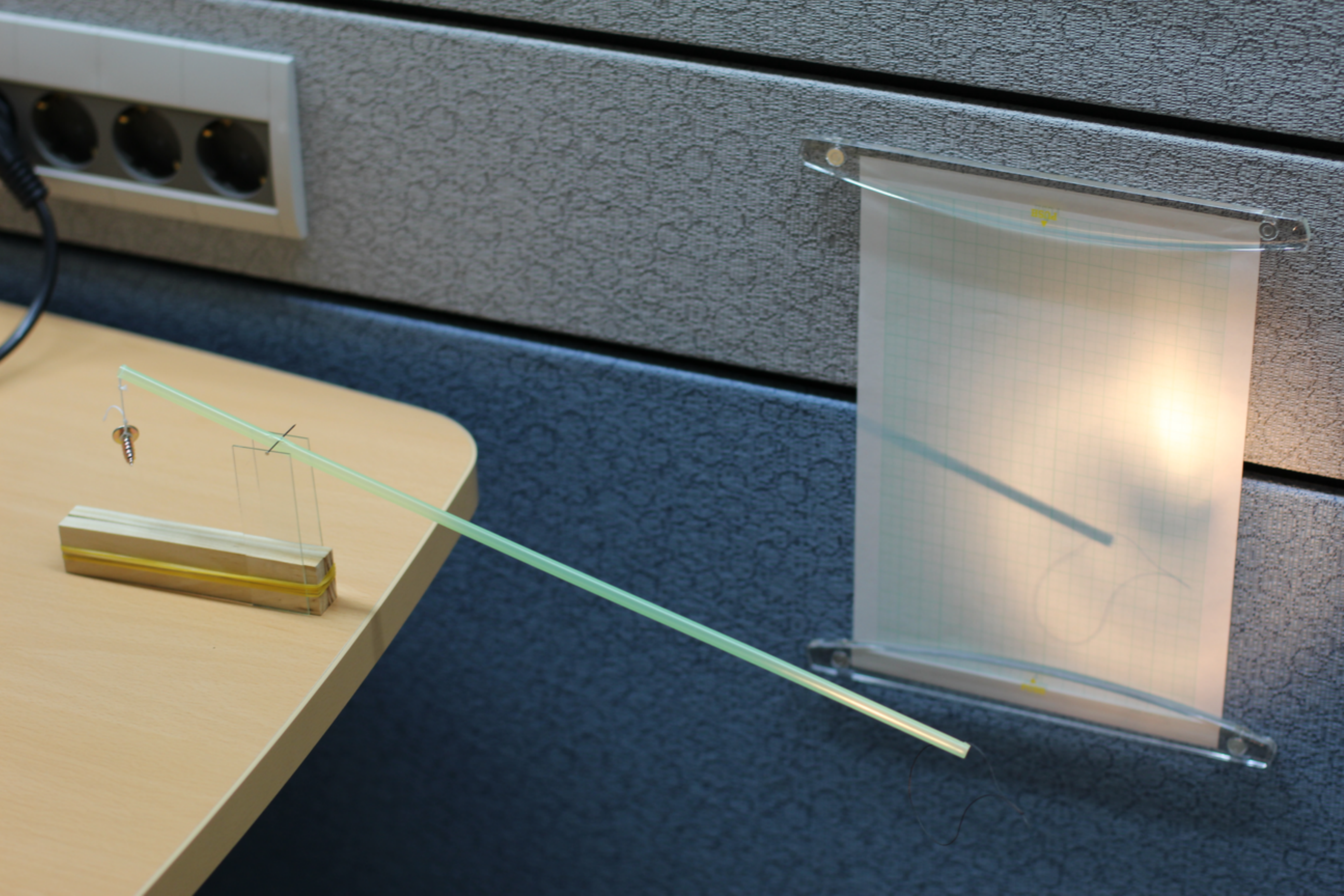}
\caption{\label{fig:expt}After placement of thread A at the other end, the balance establishes a new equilibrium with a tilted angle.}
\end{center}
\end{figure}

Figures~\ref{fig:exph} and \ref{fig:expt} show experimental setups. The pinned point is crucial. That point must be positioned upper than the center line of a straw. If you have that point under the center line you will notice that the straw does not function at all. Another consideration for a successful experiment is a sliding motion. If there is a sliding motion of the axis during placement of a test mass then the coordinate $X$ will be influenced. And this interferes with the result. Actually this is one of main sources of errors. So it is desirable to reduce the sliding motion during placement of a test material. A small scratch at the end of a straw and use of a pair of tweezers might help to place a test mass with a less introduction of the sliding. The tilted coordinates are read off from the grid paper. To mark the position of the end point of a straw we use a shadow of the straw. The measured values are summarized at Table~\ref{tab:data}. Relative errors between the measured values by a drinking straw balance and an electronic scale are order of 22\%. The uncertainties of values by a straw balance are order of 1 mg.

\section{Conclusion}
The rolling motion of an axis is a key to the understanding of working mechanism of a drinking straw balance. Due to the rolling of an axis and the relative position between the axis and the center line of a straw, ratio of lever arm lengths can change and a new equilibrium is possible. With a do-it-yourself drinking straw balance we measured masses of test materials. The data are in agreement with that of an electronic scale within 22\%.

\begin{acknowledgments}
HK is supported by Korea Science Academy of KAIST.
\end{acknowledgments}

\newpage

\begin{table}[h]
\begin{center}
\begin{tabular}{|c|c|c|c|c|c|}
\hline
\phantom{pp}Material\phantom{pp} & \phantom{p}$X$ (mm)\phantom{p} & \phantom{p}$Y$ (mm)\phantom{p} & \phantom{p}$\delta m_{\textrm{balance}}$ (mg)\phantom{p} & \phantom{pp}$\delta m_{\textrm{scale}}$ (mg)\phantom{pp} & \phantom{p}Relative error (\%)\phantom{p} \\
\hline
Thread A & \phantom{p}11 & \phantom{p}92 & \phantom{p}$14 \pm 1$ & \phantom{p}18 & $-22.2$\\
\hline
Thread B & \phantom{p}95 & 226 & \phantom{p}$53 \pm 2$ & \phantom{p}56 & \phantom{p}$-5.4$\\
\hline
Rubber & 189 & 284 & $130 \pm 1$ & 112 & $\phantom{+}16.1$\\
\hline
\end{tabular}
\caption{\label{tab:data}Comparisons of the measured mass values by a straw balance ($\delta m_{\textrm{balance}}$) and an electronic scale ($\delta m_{\textrm{scale}}$) for different test materials. $\delta m_{\textrm{balance}}$ is calculated from Eq.~\ref{eq:dmexac}. $\theta$ is determined numerically from Eqs.~(\ref{eq:XY1}) and (\ref{eq:XY2}). $M_{\rm st} = 1.953~{\rm g}$, $M_{\rm n} = 3.209~{\rm g}$, $l_L = 303~{\rm mm}$, $r = 2.8~{\rm mm}$.}
\end{center}
\end{table}

\newpage

\end{document}